\documentclass[usegraphicx,usenatbib]{mn2e}

\renewcommand\[{\begin{equation}}
\renewcommand\]{\end{equation}}

\def\Reff{$R_{\rm e}$}
\def\sp{$\,$}

\def\nh{{\cal V}}
\def\rc{r_{c}}

\catcode`\@=11
\def\gsim{\ifmmode{\mathrel{\mathpalette\@versim>}}
    \else{$\mathrel{\mathpalette\@versim>}$}\fi}
\def\lsim{\ifmmode{\mathrel{\mathpalette\@versim<}}
    \else{$\mathrel{\mathpalette\@versim<}$}\fi}
\def\@versim#1#2{\lower 2.9truept \vbox{\baselineskip 0pt \lineskip
    0.5truept \ialign{$\m@th#1\hfil##\hfil$\crcr#2\crcr\sim\crcr}}}
\catcode`\@=12

\begin{document}

   \title{Reconciling optical and X-ray mass estimates: the case of the
elliptical galaxy NGC3379}

   \author[Pellegrini \& Ciotti]
          {Silvia Pellegrini \& Luca Ciotti
           \\
		   Astronomy Department, University of Bologna, 
                       via Ranzani 1, 40127 Bologna, Italy
          }

   \maketitle

\begin{abstract} 

NGC3379 is a well studied nearby elliptical for which optical investigations
have claimed a little dark matter content, or even no dark
matter. Recently, its total mass profile $M(r)$ has been derived 
by exploiting $Chandra$
observations of its extended and X-ray emitting interstellar medium,
based on the hypothesis of hydrostatic equilibrium for the hot gas.
The resulting total mass within the effective radius \Reff\sp has
been claimed to be a few times larger than found by optical
studies. Here we show that part of the discrepancy can be due to an
underestimate of the optically derived mass, and the remaining
discrepancy of a factor of $\sim 2$ can be explained by deviations
from hydrostatic equilibrium of the hot gas. By using hydrodynamical
simulations tailored to reproduce the observed hot gas properties of NGC3379, 
and by assuming as input for the simulations
the total mass profile derived optically, we show that i) the hot gas
at the present time has X-ray properties consistent with those
observed only if it is outflowing over most of the galactic body, and
ii) an overestimate of $M$ of the same size found in the recent
X-ray analysis is recovered when assuming hydrostatic equilibrium.  We
also show that the hot gas is outflowing even for a dark matter
fraction within \Reff\sp as large as derived with the standard X-ray
procedure based on the hydrostatic equilibrium assumption, which shows
the unapplicability of the method for this galaxy. Finally, we find
that the whole range of dark mass amount and distribution allowed for
by optical studies is compatible with a hot gas flow with the observed
X-ray properties.

\end{abstract}


\section{Introduction}

The presence of an extended and X-ray emitting interstellar medium
(ISM) in early type galaxies has been often used as a tool to
determine their dynamical properties, in particular their mass profile
and total mass out to large radii ($\sim 10$ optical effective radii
\Reff; e.g., Fabbiano 1989). Most recently, this kind of analysis has
been performed for a sample of 53 early type galaxies by exploiting
$Chandra$ data (Fukazawa et al. 2006, hereafter F06).  In the commonly
adopted approach, a spherical mass distribution is assumed and, in
the hypothesis of hydrostatic equilibrium, the total mass profile is
recovered via the well known formula (Fabricant, Lecar \& Gorenstein
1980):
\[ 
M(r)=-{k
Tr\over G\mu m_p}\left( {d\,{\rm log}\rho \over d\,{\rm log} r}
+{d\,{\rm log}T \over d\,{\rm log} r} \right),  
\] 
where $T$ and $\rho$ are the radial temperature and density profiles,
inferred from deprojection techniques (e.g., White, Jones \& Forman
1997), $m_p$ is the proton mass, $\mu$ is the mean molecular weight
and $k$ is the Boltzmann constant.  Using eq. (1), several authors
measured the total mass of X-ray bright, nearby early type
galaxies, finding in general larger masses than obtained with optical
studies (e.g., Ciotti \& Pellegrini 2004 and references therein;
hereafter CP04), or sometimes lower than these, in the central
galactic regions (Tsai 1994). In fact mass profiles can be derived
also from the observed stellar velocity dispersion profiles
$\sigma_p(R)$; these extend out to $\sim 1-2$\Reff\sp in few best
cases (e.g., Bertin et al. 1994, Kronawitter et al. 2000).  For
example, by solving the Jeans equations for assumed visible and dark
density profiles and a specified radial trend of anisotropy in the
stellar velocity dispersion tensor, one obtains $\sigma_p(R)$ to be
compared with the observed one.  This method suffers from a ``dark
matter--anisotropy degeneracy'', because radial orbital anisotropy can
mimic the presence of a massive dark matter halo (e.g., Tonry
1983). This technique has also been used considering as tracers
planetary nebulae or globular clusters (Romanowsky et al. 2003,
Teodorescu et al. 2005; Pierce et al. 2006), giving a variety of
results.  However, optical studies converge towards an amount of dark
mass that is less than the optical mass within \Reff\sp [or even much
lower, see Cappellari et al. (2006) for the most recent analysis based
on SAURON data], and possibly larger outside.  The use of the hot ISM to
derive dynamical quantities using eq. (1) is then appealing both for
its capability to reach large radii and for being exempt from the
orbital anisotropy degeneracy problem.

In a recent work, however, the accuracy of the X-ray based estimates
of the total mass with eq. (1) [hereafter $M^X_{est}(r)$] has been questioned
(CP04). Instead of eq. (1), a more general
expression for the momentum equation of the hot ISM is (e.g., Ciotti
et al. 1991):
\[
{k\over \mu m_p \rho} {d\, \rho T\over dr}+\nh=- g, 
\] 
where $g=GM(r)/r^2$ describes the total gravitational field and $\nh$
is the kinetic term that includes all the non hydrostatic contributions, such as
the time dependence of the velocity, the advective term and the presence 
of mass sources. CP04 investigated the
effect that deviations from hydrostatic equilibrium have on $M^X_{est}(r)$
derived via eq. (1), and showed that
\[
M^{X}_{est}=M+{r^2 \over G}\nh.
\]
They found that less massive distributions than true are deduced for the
central regions of galaxies hosting inflowing gas, and conversely that the
true mass is overestimated for outflowing regions where the velocity increases
with radius, as in galactic winds.

Currently, quite accurate $T$ and $\rho$ profiles are being derived
from $Chandra$ and $XMM-Newton$ observations (e.g., Sun et al. 2003;
F06). F06, in particular, investigated the dynamical structure of 53
elliptical galaxies, both X-ray bright and X-ray faint; by exploiting
the high angular resolution of $Chandra$, they obtained a temperature
and density profile from 0.1 to 10 \Reff.  For the well studied nearby
elliptical NGC3379, they derived an $M^X_{est}$ of $\sim 7$ times the
value obtained with optical methods, for the same galactic region.
Here we demonstrate how deviations from hydrostatic
equilibrium can be responsible for the discrepancy between
$M^{X}_{est}$ and the mass estimated from optical studies, for this
galaxy. By using hydrodynamical simulations tailored to reproduce the
hot gas properties observed by $Chandra$ for NGC3379, it is shown that
an overestimate is obtained from the use of eq. (1), since the hot ISM
is largely outflowing. In Section 2 we briefly describe the previous
mass estimates for NGC3379 coming from optical studies and from the
recent X-ray analysis; in Section 3 we present the galaxy model used
for the hydrodynamical simulations and the results of the simulations;
in Section 4 the conclusions are presented.

\begin{table*}
\begin{center}
\caption{General and hot ISM properties of NGC3379}
\begin{tabular}{cccccccccccccc} 
\hline\hline 
   D  & Morph. & $B^T_0$& $L_B$                & \Reff$^a$              & $\sigma_c^b$ &    $ L_X ^c$ &   band & $kT^c$  \\
 (Mpc)& type   & (Bmag) &$(10^{10}L_{B,\odot}$)&($^{\prime\prime}$, kpc)& (km s$^{-1}$)& (erg s$^{-1}$)& (keV) & (keV) \\
\hline
 10.6 & E1     & 10.18  & 1.5                  & 37.5, 1.93             &    221       &4.6$\times10^{37}$&0.2--5&0.52$\pm0.30$\\
\hline \hline							
\end{tabular}							
\end{center}

D is from Tonry et al. 2001; the morphological Type and $B^T_0$ are
from RC3; $L_B$ derives from $B^T_0$ and the 
adopted distance.

$^a$ Peletier et al. (2000)

$^b$ central stellar velocity dispersion from Hallyday et al. (2001)

$^c$ Fukazawa et al. (2006)

\end{table*}

\section{The total mass profile of NGC3379}

NGC 3379 is the dominant elliptical galaxy in the nearby Leo Group,
and has an optical profile well described by a de Vaucouleurs (1948)
law over
a span of ~10 magnitudes (Capaccioli et al. 1990; see Table 1 for the
main galaxy properties). Its optical shape is roundish (its
morphological type is E1) and its optical colors are typical of an old
population; it shows no sign of current or past interactions
(Schweizer \& Seitzer 1992).
The total mass profile of NGC3379 has been studied several times using
optical data (hereafter, all quantities involving distance have been 
rescaled for the distance in Table 1; $M$ indicates the total mass, 
$M_*$ the stellar mass and $M_h$ the dark halo mass).

\subsection{Results of optical studies}

Even when performed with different approaches, optical studies
invariably show very little dark matter within \Reff. Their results
are usually expressed in terms of the B-band total mass-to-light ratio
$M/L_B$.  For NGC3379 the contribution of $M_*/L_B$ to the total
$M/L_B$ has been estimated to range from 4.2 to 8.8 from stellar
population synthesis models (Gerhard et al.  2001).

The dynamical modeling by Saglia, Bertin \& Stiavelli (1992), based on
kinematical data extending out to $\sim 1.7$\Reff, gives a best fit
$M/L_B=7.8$ for a one-component modeling, for the whole galaxy.  In
their phase-space based two-component modeling, they found 
$M_*/L_B= 7.5$, and a dark-to-luminous mass ratio
$M_h/M_*=0.67^{+0.16}_{-0.30}$ within \Reff,
and $M_h/M_*=1.4^{+3.6}_{-0.7}$ for the whole galaxy.

Kronawitter et al. (2000)'s dynamical modeling, based on stellar
kinematic information including line profile, and extending from 0.005 to
2.75\Reff, shows an outward increase of $M/L_B$ of a factor of
$1.2^{+0.4}_{-0.2}$, with respect to a central value of
$M/L_B=5.6_{-0.37} ^{+0.25}$ (though the modeling is also consistent
with a flat $M/L_B$ over the range of the kinematic data).  At \Reff\sp
they give\footnote{This is a ratio between spatial quantities, that 
is between the mass and
luminosity enclosed within a sphere of radius $r=$\Reff\sp (Saglia, private
communication).} $M/L_B=6.3$, which implies $M_h/M_*= 0.13$ when for
$M_*/L_B$ the central $M/L_B$ is adopted. At their
outermost datapoint, $M_h/M_*=0.2^{+0.4}_{-0.2}$. 

More recent modeling of long-slit data extending to
80$^{\prime\prime}$ found again that dark matter is not required out
to $\sim 2.1$\Reff\sp (Samurovic \& Danziger 2005); the data can be
fitted with a constant $M/L_B=6-11$.  SAURON observations of the
central region of NGC3379 give integral-field stellar kinematics out
to $28^{\prime\prime}$ (Cappellari et al. 2006). Coupled with
2-integral Jeans and 3-integral Schwarzschild dynamical models, the
data are consistent with a constant mass to light ratio in the $I$
band; from their $M_*/L_I$ for the stellar population, $M_h/M_* \sim
0.09$ is recovered.

The mass profile has been studied also at larger radial distances
using planetary nebulae as tracers.  Ciardullo, Jacoby, \& Dejonghe
(1993) presented a radial velocity survey of planetary nebulae with
galactocentric radii $\lsim 4.5$\Reff. At large radii, the stellar
dispersion profile joins with that of the planetary nebulae.  The
velocity dispersion and photometric profiles agree with a constant
mass-to-light ratio isotropic orbit model with $M/L_B\sim 7$, similar
to what is obtained with anisotropic models; therefore it was
concluded that dark matter is not needed within $\sim 4.5$\Reff.

Romanowsky et al. (2003) found a decrease in the velocity dispersion
profile of planetary nebulae extending out to $200^{\prime\prime}$.
Within this radius, their isotropic modeling constrained $M/L_B=7.1\pm
0.6$, a result consistent with no dark matter, given the possible
$M_*/L_B$ values (Gerhard et al. 2001).
Dekel et al. (2005) questioned this result with radially
anisotropic modeling, and showed that a total $M/L_B\sim 15$ at
5\Reff\sp is consistent with the data; this implies $M_h/M_*=1.1-1.5$ 
(for an assumed $M_*/L_B=6-7$).

Summarizing, the optical studies converge towards a very small dark
matter content within \Reff, of the order of $M_h/M_*\sim 0.1$,
except for the Saglia et al. (1992)'s analysis that gives $M_h/M_*
=0.37-0.83$. At larger radii, the dynamical modeling gives results
consistent with no dark matter, but also with a global galactic value
of $M_h/M_*=1.1-1.5$ (Dekel et al. 2005) or even $M_h/M_*=0.7-5$
(Saglia et al. 1992).

\subsection{Results of X-ray studies and first 
reduction of the discrepancy}

The hot gas content of NGC3379 is very low, as known since $ROSAT$
observations (e.g., Pellegrini 2005, Brown \& Bregman 2001, Trinchieri et
al. 2000).  The most recent $Chandra$ data have been used by F06 to
derive the total mass within \Reff.
From a parameterized gas density and temperature profile entered in
eq. (1), F06 obtained $M^X_{est}=8.7\times 10^{10}M_{\odot}$, with a
quoted 'typical error' of 30\% [therefore $M^X_{est}=(8.7\pm
2.6)\times 10^{10}M_{\odot}$].  From within the same region, we derive
a total $M=3.7\times 10^{10}\, M_{\odot}$, by taking $M/L_B$ at
\Reff\sp of Kronawitter et al. (2000) (see Sect. 2.1) and\footnote{For
the de Vaucouleurs (1948) law, the luminosity enclosed within a
spherical radius $r=$ \Reff\sp is $\simeq 0.42$ of the total one.}
0.42$\times $ their quoted $L_B$.  Therefore, we determine a discrepancy of
a factor of $\sim 2$ between the X-ray and the optical estimate.

F06 noted a larger discrepancy with the mass derived by
Kronawitter et al. (2000), by a factor of $\sim7$  (see
their Table 5); however, they
likely underestimated the Kronawitter et al.'s total mass within
\Reff, for which they give just 1.3$\times 10^{10}M_{\odot}$. In fact,
this mass cannot be smaller than $(M/L_B)_c \times 0.42 L_B =
3.3\times 10^{10} M_{\odot}$, where $(M/L_B)_c=5.6$ is the central
mass-to-light ratio of Kronawitter et al. In any case, $M_*/L_B>4$ for
the old stellar population of NGC3379 (Gerhard et al. 2001), which
already implies $M_*>2.5\times 10^{10} M_{\odot}$ within $r=$ \Reff.

All in all, the observational results
that will be considered in the following are that i) $M^X_{est}$
within \Reff\sp is a factor of $\sim 2$ larger than the total mass
estimated by Kronawitter et al. within the same radius; ii) when
taking an average $M_*/L_B\sim 6$ for the stellar population and the
$L_B$ value in Table 1, $M_{est}^X$ corresponds to $M_h/M_*
\sim 1.3$ within \Reff.

In the next Section, with the aid of hydrodynamical simulations, we
show how an overestimate of a factor of $1.5-2$ for the total mass of NGC3379
can be easily obtained when using eq.(1) instead of eq. (2),
since the hot gas is not in hydrostatic equilibrium, but instead it is
outflowing over most of the galactic body.  A summary of the hot gas
properties derived by F06 is given in Table 1; even though resolved
point sources have been subtracted before deriving these properties,
some residual emission from unresolved stellar sources is still
contributing to the X-ray luminosity of the gas in Table 1.

\section{The hydrodynamical simulations}

The time-dependent equations of hydrodynamics with source terms, and
the numerical code used to solve them, are fully described in Ciotti
et al.  (1991). The model galaxy is initially devoid of gas, a
situation produced by the galactic winds established by type II
supernovae, early in the evolution of elliptical galaxies. Then the
mass losses from the evolving stellar population start replenishing
the interstellar space, and the resulting gas
flow evolution is followed until an age of $\sim 9$ Gyrs is reached,
that is the estimated age for NGC3379 (Terlevich \& Forbes 2002 give
an age of $9.0^{+2.3}_{-1.9}$ Gyr).

The X-ray luminosity in the energy band adopted by F06 has been
calculated using the APEC code in XSPEC, for the density and
temperature distribution of the gas given by the simulations (solar
abundance has been adopted). The luminosity weighted emission
temperature in the same energy band has also been calculated for the
models.

\subsection{The galaxy model}

A very good approximation of the de Vaucouleurs (1948) law is given by the
Hernquist (1990) distribution, that has the advantage that all its
dynamical properties can be expressed analytically.
This distribution is

\begin{equation}
\rho (r)={M\over 2\pi} {\rc\over r (\rc+r)^3}
\end{equation}
where $\rc$ is the scale length and relates to the effective radius of
the de Vaucouleurs law as $\rc \simeq$\Reff/1.8153. This law gives a 
description of the luminous mass distribution
accurate enough for the treatment of hot
gas flows, on the scales that are relevant for the problem (from few
tens to several thousands of parsecs). In the simulations
we adopt a central grid spacing of 5 pc to allow
for a better sampling of the inner regions. 

The radial density distribution of the dark haloes of ellipticals is
not well constrained by observations; theoretical arguments favor a
peaked profile (Ciotti \& Pellegrini 1992; Evans \& Collett 1997), and
high resolution numerical simulations of dissipationless collapse
produce a density distribution $\propto r^{-1}$ near the center
(Dubinsky \& Carlberg 1991, Navarro, Frenk, \& White 1996).  The model
galaxy is therefore a superposition of two Hernquist density
distributions, one for the luminous matter $\rho _* (r)$ with
$r_c=r_{*}$ and $M=M_*$ in eq. (4), and one for the dark matter $\rho
_h (r)$ with $M=M_h$, $r_c= r_{h}=\beta r_{*}$.

In order to build a galaxy model where the masses of the two
components and their distributions are consistent with observations
and also produce the observed $\sigma_c$, we solve and project the
Jeans equations for a globally isotropic two-component Hernquist
model.  For this model we find that a very accurate fit of
the projected, mass averaged stellar velocity dispersion within \Reff/8 is
\begin{equation} 
\sigma ^2_c\simeq {GM_*\over r_*} \times \left (
0.096+0.12 {M_h/M_* \over \beta^{1.72}} \right ),
\end{equation} 
with
\begin{equation} 
\beta=2.8153 \sqrt{ M_h/M_* \over (M_h/M_*)_{\rm e} }-1.8153,
\end{equation} 
where $ (M_h/M_*)_{\rm e}$ is the dark-to-luminous mass ratio within a
radius $r=$ \Reff.  Thus for assigned \Reff, $\sigma_c$, total
$M_h/M_*$ and $ (M_h/M_*)_{\rm e}$, we derive $M_*$ and $\beta$ from
eqs. (5) and (6).  The typical stellar mass-to-light ratios resulting
from this procedure, for $(M_h/M_*)_{\rm e}=0.1-1$, are between 4 and
8. The model phase-space distribution function is everywhere positive
(Ciotti 1996).

\subsection{Time evolving inputs}

The time evolving input ingredients of the numerical simulations are
the rates of stellar mass loss from the aging stellar population, and
the rate of SNIa heating.  In the numerical code the exact mass return
rate prescribed by the stellar evolution theory is used (see, e.g.,
Ciotti et al. 1991), updated to take into account the latest stellar
population synthesis models (Maraston 2005) and more recent estimates
for the mass ejection from stars as a function of their mass (Renzini
\& Ciotti 1993). The SNIa heating rate is parameterized as $L_{SN}(t)=
E_{SN}\, R_{SN}(t)\,L_B$, where $E_{SN}=10^{51}$ erg is the kinetic
energy injected in the ISM by one SNIa, and the number of events as a
function of time is $R_{SN}(t)L_B\propto t^{-s}$, where the slope $s$
describes the unknown decay rate (see also Pellegrini \& Ciotti 1998).
$R_{SN}(t)$ is normalized to give the SNIa's explosion rate in nearby
E/S0s for $t=10$ Gyrs, i.e., 0.18$h_{75}^2$ SNu (where 1 SNu = 1 SNIa
per 100 yrs per $10^{10}L_{B,\odot}$, $h_{75}=H_{\circ}/75$), that has
an associated uncertainty of $\sim 30$\% (Cappellaro et al. 1999).
Therefore we consider as a free parameter also the ratio $\vartheta$
between the adopted SNIa's rate and the best fit estimate of
Cappellaro et al., with $\vartheta=0.67-1.33$.  From a comparison
between realistic modelings of the rate and its observation in
high-$z$ surveys currently available, a slope $s\sim 1.0-1.2$ has been
favoured (Greggio 2005).

Another heating source, though less important than SNIa's, is the
thermalization of the stellar mass losses to the local "temperature"
set by the stellar velocity dispersion. At each radius this is determined
using the velocity dispersion profile obtained by solving the Jeans
equation for the two-component Hernquist model in the globally
isotropic case (Ciotti 1996).  If anisotropy is present, this heating
differs locally from the value it has in the isotropic case. However,
the difference is small for realistic anisotropy distributions, and
the total heating over the galaxy is independent of orbital anisotropy,
as a consequence of the virial theorem. In addition,  the heating due to
thermalization of stellar motions is significantly smaller than SNIa heating,
and therefore the effects of anisotropy on $M^X_{est}(r)$ are negligible
[they are just indirect effects in eqs. (7)-(8) of the next Section, through 
the gas temperature and velocity terms].

\begin{figure}
\vskip -0.7truecm
\hskip -1.7truecm
\includegraphics[height=.58\textheight,width=.68\textwidth]{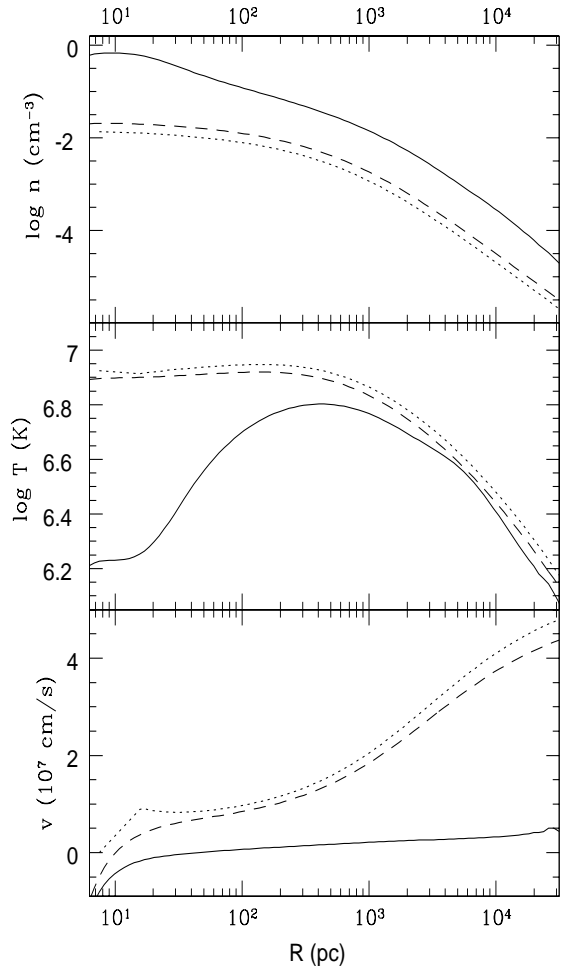}
\caption{From top to bottom the radial profiles of the gas density, 
temperature and velocity for representative models of inflow (solid line), 
partial wind (dashed line) and global wind (dotted line);
see Sect. 3.3.}
\end{figure}

\subsection{Results}

In Sect. 3.3.1 below we adopt the underlying total mass distribution on
which most of the modeling of the optical data converge (Sect. 2.1).  In
Sect. 3.3.2 we use as input the mass model found by F06 and also a
"maximum" dark matter model. In both Sections we vary the SNIa's
parameters $\vartheta$ and $s$ to obtain X-ray gas properties as
observed when an age of $\sim 9$ Gyrs is reached. When this is the
case, we apply to
the radial distributions of the gas density and internal energy,
resulting from the numerical simulation, the standard procedure of
eq. (1), in order to recover $M^X_{est}(r)$ and compare it with the true
underlying mass distribution $M(r)$.  In practice, we calculate the
"empirical" $M_{est}^X(r)/M(r)$ ratio from the output of the numerical
simulations as
\[ 
{M_{est}^X(r)\over M(r)} = -{k\over
\mu m_p \rho g} {d\,\rho T \over d\, r}.
\]
This ratio is also given by eq. (3), where the $\cal V$ term can be
explicitely calculated from the momentum equation [see eq. (10) of
CP04)].  In quasi-stationary conditions, as those reached by our
models at the end of the simulations, one can neglect the $\partial
u/\partial t$ term in the expression for $\cal V$, and the full
expression for eq. (3) becomes:
\[ 
{M_{est}^X(r)\over M(r)} = 1+{ u \dot\rho \over \rho g} +
{1\over 2 g} {d\,u^2 \over d\, r}
\] 
where $\dot \rho$ is the local mass injection rate due to stellar mass losses
and $u$ is the radial flow velocity.
Eq. (8) allows for a quantitative evaluation of the separate contributions
of the non hydrostatic terms to the $M_{est}^X(r)/M(r)$ ratio.

By varying the input parameters, the resulting gas flow phases are of
a global wind, a partial wind (where the gas is inflowing only over a
very small central region) and a global inflow; a sample of the radial
profiles of gas velocity, temperature and density for 
representative models of these three phases at the present epoch is
shown in Fig. 1 [see Pellegrini \& Ciotti (1998) for a more detailed 
description of the gas flow evolution].

\subsubsection{Optically derived total mass model}

In this experiment the galaxy model is based on the conclusions of Sect. 2.1:
it has $(M_h/M_*)_{\rm e}=0.13$, in accordance with
Kronawitter et al. (2000), and a global $M_h/M_*=1.5$, consistent with
the results of Dekel et al. (2005) and Saglia et al. (1992).  
The other parameters are set to the observed values given in Table 1.

With this total mass profile, the hot ISM maintains in a global wind
phase throughout its evolution, if the SNIa's rate is such that
$\vartheta =1$, regardless of the slope $s$. After a time equal to the
age estimated for NGC3379, the hot ISM luminosity in the
0.2--5 keV band is less than derived by F06 by at least an order of
magnitude. If the SNIa's rate is lowered, the hot gas flow maintains
in a wind phase over most of the galactic body, while it develops an
inflow in a very small central region ($\lsim 20$ pc; see the partial
wind case in Fig.1). In this way a higher luminosity is obtained at
the present epoch, that is comparable to the observed one when
$\vartheta$ is close to its lowest values ($\vartheta \sim 0.7$).

In Fig. 2 the derived $M_{est}^X/M$ is shown for a sample case of a
partial wind with a hot gas luminosity of $L(0.2-5$ keV)$=3.6\times
10^{37}$ erg s$^{-1}$ and an emission weighted temperature of 0.53
keV, both values close to those reported by F06 (see Table 1)\footnote{A
contribution from unresolved sources to L(0.2--5 keV) in Table 1
(Sect. 2.2),
and the uncertainty associated with this measurement, can 
account for the slightly lower L(0.2--5 keV) of the model.}.
It is obtained for $s=1.1$ and $\vartheta=0.72$, after a time of 8 Gyrs.
$M_{est}^X(r)$ overestimates $M(r)$, by a factor of $\sim 1.5-1.6$
around \Reff. This factor is in the direction of, and also enough for,
the explanation for the residual discrepancy between $M^X_{est}$
derived by F06 and the total mass derived with optical methods
(Sect. 2.2).  Note how the $M_{est}^X/M$ profiles derived with eq. (7)
and with eq. (8) maintain remarkably close to each other over the
whole numerical grid. This means that the terms describing the
deviation from hydrostatic equilibrium and entering eq. (8) are really
accounting for the whole discrepancy measured with eq. (7) (and it is 
also an independent check of the reliability and accuracy of the 
numerical code).

\begin{figure}
\vskip -6.8truecm
\hskip -1.3truecm
\includegraphics[height=.65\textheight,width=.7\textwidth]{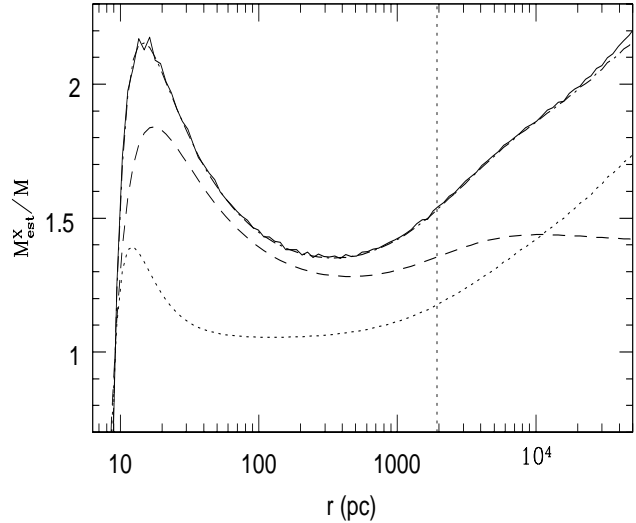}
\caption{The radial trend of $M^X_{est}/M $ obtained from eq. (7), for the
partial wind model described in Sect. 3.3.1 (solid line). Also shown are 
the terms on the right hand side
of eq. (8): the radial derivative term (plus unity, dotted line), 
the term with  the mass source (plus unity, dashed line), and 
$M^X_{est}/M $ according to eq. (8)
(dot-dashed line). The dotted vertical line marks the position of
\Reff.  }
\end{figure}

\begin{figure}
\vskip -6.8truecm
\hskip -1.3truecm
\includegraphics[height=.65\textheight,width=.7\textwidth]{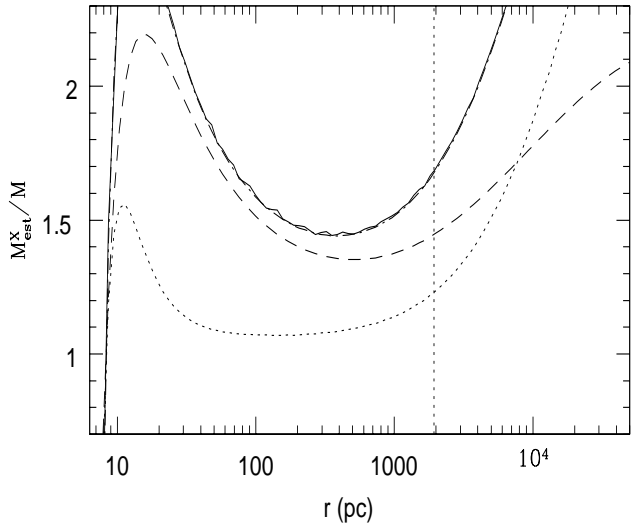}
\caption{The same as in Fig. 2, for a model with a larger fraction of
dark matter content within \Reff\sp [$(M_h/M_*)_{\rm e}=1.3$], consistent
with the X-ray analysis of F06 (see Sect. 3.3.2). Again
$\vartheta =0.72$ and $s=1.1$.}
\end{figure}

\subsubsection{X-ray derived and maximum dark mass model}

We study here a galaxy model with $(M_h/M_*)_{\rm e}=1.3$, in
accordance with the $M^X_{est}$ derived by F06 (Sect. 2.2); the global
$M_h/M_*=1.5$ as in the previous Section 3.3.1, and the SNIa's
parameters are varied.

Again, for $\vartheta=1$ global winds that are far less luminous than
the observed $L_X$ are obtained, when the galaxy is 9 Gyrs old.
Luminosity and temperature close to the observed values are shown
again when the flow is a partial wind, with approximately the same low
SNIa's rates derived in the previous Section ($\vartheta \sim 0.7$).
This fact can be explained by considering that, although
$(M_h/M_*)_{\rm e}$ increases from 0.13 to 1.3, the total mass in the
central region remains the same, and the absolute value of the binding
energy per unit mass decreases by $\sim 15$\% at the center;
this is the outcome of imposing fixed values for the central stellar
velocity dispersion and \Reff\sp (see eqs. 5 and 6). In addition,
being $L_B$ also fixed, the rate of mass and heat input are the same
as in the set of models of Sect. 3.3.1. All in all, the net result is
just a small reduction of the binding energy in the central region,
where almost all of the emission is produced.

Since the flow pattern is the same as in the previous Section, also
the mass estimates $M^X_{est}$ suffer from the same problems. Fig. 3
shows such an estimate for a model with $L_X=2.5\times 10^{37}$ erg
s$^{-1}$ and a temperature of 0.53 keV, after 8 Gyrs; the input SNIa's
parameters are $s=1.1$ and $\vartheta=0.72$. While similar to Fig. 2
in the general trends of the various quantites, the overestimate is
even larger than in Fig. 2.  Thus, even if $(M_h/M_*)_{\rm e}$ were
really as high as deduced from assuming hydrostatic equilibrium, the
derived $M_{est}^X$ would be wrong.  This experiment also shows that
the hot ISM is quite insensitive to the underlying mass distribution
and partition between stellar and dark mass within \Reff, at least for
these low dark matter contents.

In the last experiment, we check whether reliable mass
estimates can ever be obtained with the X-ray procedure of eq. (1),
for this galaxy.
This requires the hot gas to be roughly in hydrostatic equilibrium
over most of the galaxy, as in a slow global inflow (e.g., Ciotti et
al. 1991). In order to obtain an inflow over a large region, the dark
matter content must be increased. Therefore, we assume the maximum
dark matter content allowed for by optical studies (Saglia et
al. 1992), that is a global ratio of $M_h/M_*=5$ and $(M_h/M_*)_{\rm
e}=0.83$ (note that this latter value is also consistent with the 
estimate of F06, within the uncertainties; see Sect. 2.2).
If $\vartheta \sim 0.7$, the galaxy indeed hosts a global inflow;
however, after
an age of 9 Gyrs, it is far more luminous than observed ($L_X\gsim
10^{40}$ erg s$^{-1}$).  For $\vartheta > 0.7$, the flow pattern is
again a global inflow during the first part of the evolution, but
before reaching an age of 9 Gyrs, it slows down until it reverts its
direction into an outflow, over most of the galaxy. The final phase is
a partial wind, whose luminosity is much lower than in a model where
the gas is still mostly inflowing. For example, if $\vartheta=0.92$
and $s=1.0$, the transition epoch is at $\sim 9$ Gyrs, after which the
luminosity decreases by 2 orders of magnitude in $\lsim 1$ Gyr.  The
observed L(0.2--5 keV) is reached between 10 and 10.5 Gyrs, with an
average emission temperature of $\sim 0.4$ keV.  Fig. 4 shows the mass
profile that would be derived for this model: as for the previous
cases, an overestimate is obtained at \Reff.  This experiment shows
that the X-ray procedure applied to NGC3379 is forced to produce mass
overestimates, because the low observed X-ray emission requires that
the gas is outflowing over most of the galaxy.

\begin{figure}
\vskip -6.8truecm
\hskip -1.3truecm
\includegraphics[height=.65\textheight,width=.7\textwidth]{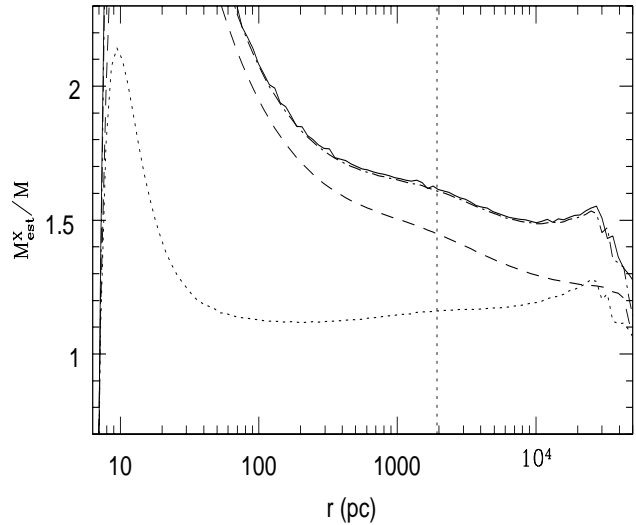}
\caption{The same as in Figs. 2 and 3, for the maximum
dark mass model, where an inflow reverts into a PW at 
$\sim 9$ Gyrs (see Sect. 3.3.2).}
\end{figure}

\section{Conclusions}

We have analyzed the discrepancy between optical and X-ray mass
estimates of NGC3379, a well studied nearby elliptical for which
optical investigations are consistent even with absence of dark
matter, but for which recent X-ray studies derived a total mass much
larger than derived with optical methods, within \Reff.
First we have reduced the discrepancy to a more plausible factor of
$\sim 2$, by a reanalysis of the total mass within \Reff\sp implied by
previous optical studies. Then we have argued in favour of the
hypothesis of lack of hydrostatic equilibrium as the explanation for
the residual inconsistency between the X-ray and optically derived
total mass.

With the aid of hydrodynamical simulations tailored specifically for
NGC3379, we have tested whether hot gas properties similar to those
observed can be recovered, and then what total mass is derived if the
standard X-ray procedure is applied to the flow resulting from the
simulation. In a first experiment the total mass distribution
underlying the gas flow is that derived by the modeling of the optical
data, with $(M_h/M_*)_{\rm e}=0.13$.  For a SNIa's explosion rate
consistent with current estimates, a hot gas flow with
luminosity and temperature close to the $Chandra$ values
can be obtained. However, except for a very small central
region, the gas is largely outflowing, with a velocity increasing
outwards. For such a flow, the X-ray method based on the hypothesis of
hydrostatic equilibrium gives an overestimate of the total mass by a
factor of $\sim 1.6$, within \Reff, and by $\gsim 2$ at larger radii.

In a second experiment, the underlying total mass distribution is that
derived by the modeling of the X-ray data within \Reff, under the
assumption of hydrostatic equilibrium; i.e., $(M_h/M_*)_{\rm e}$ is
increased to 1.3, while the global $M_h/M_*$ is the same as before.
Again, the observed hot gas properties are reproduced by a partial
wind; therefore, the assumption of hydrostatic equilibrium and the use
of the X-ray method produce a mass overestimate of a similar factor, 
showing the
inconsistency of the whole procedure for this galaxy.  This experiment
also shows that the hot ISM is quite insensitive to the underlying
distribution of dark mass within \Reff, at least for these low dark
matter contents.

Finally, we have tried a "maximum" dark matter model, where the amount
of dark mass is the maximum allowed for by optical studies [$M_h/M_*=5$ and
$(M_h/M_*)_{\rm e}=0.83$, which is also consistent with the X-ray
results]. The flow can now keep itself in a global inflow, the gas is
close to hydrostatic equilibrium over most of the galaxy, and the
galaxy mass estimated with the X-ray method would be reliable, but the
X-ray luminosity is orders of magnitude higher than observed.  A
luminosity as low as observed is obtained if the SNIa's rate is
increased, so that the flow starts as a global inflow and becomes a
partial wind in later times.  Therefore the X-ray method again
produces an overestimate of the total mass at \Reff.

In conclusion, NGC3379 is a clear example where the departure from
hydrostatic equilibrium of the hot ISM gives a consistent explanation
for the discrepancy between the optical and X-ray--derived mass
values. In particular, this departure (while also required to
reproduce the observed low X-ray emission) accounts for $M^{X}_{est}$
being an overestimate and also for the {\it size} of the overestimate.
This analysis suggests that a previous understanding of the 
hot gas behavior is required before attempting dynamical
estimates with hot gas-based studies. Here, for example, the very low
hot gas content suggests by itself that most of the galaxy likely
hosts a wind. 

Lastly, we can check whether the X-ray properties of NGC3379 can put
any constraint in the controversy concerning the "dearth of dark
matter in elliptical galaxies", for which NGC3379 has been suggested
to provide an example (e.g., Romanowsky et al.  2003), later
criticized by Dekel et al. (2005). We find that hot gas with an
emission and average temperature as observed by $Chandra$ can be
present for the whole range of dark mass amount and distribution
suggested by optical works, just varying the SNIa's rate within its
estimated uncertainty (i.e., it has to be not so high to sustain a
global wind, for the lowest $M_h/M_*$ values, and not so low that a
global inflow lasts until the present epoch, for higher $M_h/M_*$
up to 5). Even higher dark matter contents than found by optical works
can actually still correspond to partial winds with the required X-ray
properties.  This insensitiveness to the underlying mass is not
totally new: although for much X-ray brighter galaxies, mass models
constrained by extended stellar dynamical data, and including
different amounts of dark matter, were found to produce similar
temperature and X-ray brightness profiles, within the uncertainties in
the X-ray measurements (Bertin et al. 1993). The situation may have improved
with the $Chandra$ and $XMM-$Newton data, at least for the X-ray brightest
galaxies.


\section*{Acknowledgments}
We thank the anonymous referee for comments that improved the presentation.
We acknowledge support from MIUR (Cofin 2004) and from contract ASI-INAF
I/023/05/0.

\end{document}